\documentclass[epj]{svjour}

\newcommand{\bqn}{\begin{equation}}
\newcommand{\eqn}{\end{equation}}
\newcommand{\bqna}{\begin{eqnarray}}
\newcommand{\eqna}{\end{eqnarray}}
\newcommand{\bary}{\begin{array}{clcr}}
\newcommand{\eary}{\end{array}}
\usepackage[dvips,dvipdfm]{graphicx}
\usepackage{color}
\usepackage{amsfonts}
\usepackage{float}
\usepackage{bm}
\usepackage{setspace}
\usepackage{mathtools}
\usepackage{amsmath}

\begin{document}

\title{Quantum interference effects in a $\Lambda$-type atom interacting with two short laser pulse trains}
\author{Gabriela Buica\thanks{e-mail:buica@spacescience.ro}}
\institute{Institute of Space Sciences, Bucharest-M\u{a}gurele, P.O. Box MG-23, R-77125, Romania}

\abstract{
We study the quantum interference between the excitation pathways in
a  three-level $\Lambda$-type atom  interacting with two short laser pulse trains
 under the conditions of electromagnetically induced transparency.
The  probability amplitude equations which describe the interaction of a three-level
$\Lambda$-type atom with two  laser pulse trains are numerically solved.
We derive analytical expressions for the population of the upper excited state
 for  resonant laser pulse trains with a rectangular temporal profile.
By varying the parameters of the laser pulse trains such as area of a single pulse,
detuning, repetition period, and number of individual pulses,  we analyze the
 quantum interference between the excitation pathways in terms of the upper excited state population.
 }
\maketitle

\date{\today}

\section{Introduction}
Due to the important development of the laser sources, quantum control
 (sometimes referred to as optimal control or coherent control) has become of
 increasing interest in the last years in both physical and chemical sciences.
 Quantum control aims at manipulating the course of the quantum dynamics of atomic or
molecular systems and is based on  quantum interferences.
Up to now three methods have been proposed in order to achieve quantum control:
 the temporal coherent control that uses the temporal quantum interference of laser fields,
optimal control that uses the techniques of pulse shaping,
and  adiabatic passage in order to realize a complete population transfer between two
 quantum states \cite{Shapiro2000,Ehlotzky}.

Through the quantum interference between two different excitation pathways the optical
response of an atomic medium could be modified and an opaque optical medium can be
 rendered  transparent to a probe field by applying an intense coupling laser field
 \cite{Harris,Marangos,Bergmann}.
This effect is named electromagnetically induced transparency (EIT) and represents a
coherent optical nonlinearity which renders a medium transparent over a narrow spectral
range within an absorption line. The absorption peak is split and both  absorption and  refraction index vanish at resonance.
The EIT was theoretically proposed by Kocharovskaya et al.  \cite{Olga} and then it was
 experimentally observed in Sr \cite{Boller}.
Since its observation, EIT  has been intensively studied and developed in the domain of
 lasing without inversion, nonlinear optics, sub-fs pulse generation, atomic coherence control,
 slow light, giant nonlinear effect, storage of light, etc. \cite{Fleischhauer}.
The basis of the EIT phenomena resides in the coherent population trapping (CPT) \cite{Bergmann}
that was for the first time discovered by Alzetta et al.  \cite{Alzetta} in the D lines of the Na atom.
Thus for a three-level $\Lambda$-type atom  interacting with a probe and a coupling laser,
 the population is trapped in the lower states and the population of the upper excited state is negligible.
In terms of quantum interference, the contributions to the probability amplitude of the
upper state from the two lower states interfere destructively and the population of the
upper state could become negligible and therefore the laser radiation is not absorbed.
This quantum interference can be created among different quantum paths reaching the
same final state with single laser pulses or by several consecutive mutually coherent
laser pulses: namely a \textit{laser pulse train}.
By using a laser pulse train ultrahigh resolution spectroscopy, with the comb linewidths
much smaller than the bandwidth of the pulse, can be obtained.  Lately, it has been shown
 that the \textit{coherent accumulation effects} of populations and atomic coherences play
a role in the coherent control of atomic or molecular  systems  with  laser pulse trains \cite{Marian,Stowe}.
Most of the latest studies regarding the coherent accumulation effects in a three-level
 atom excited by a fs laser pulse train are done either in the simplified hypothesis
of a stationary state \cite{Felinto2003} or in the weak field limit,
within the lowest order perturbation theory \cite{Felinto2004}.
By properly adjusting the amplitudes and phases of the pulses, the coherent excitation
 of a two-level system  with a train of ultrashort laser pulses reproduces the
effect normally achieved with a single frequency-chirped pulse \cite{Shapiro2008}.
The effect of the pulse area and the number of pulses of a train of ultrashort pulses was
also investigated for a two-level atom \cite{Nakajima2008}.
Recently, EIT and Autler-Townes splitting  were studied for an ultrashort train pulse
interacting with a $\Lambda$-type atom driven by a continuous wave (CW) laser \cite{Soares2010}.
The coherent population trapping has been investigated with
 a train of ultrashort pulses of rectangular pulse shape in the steady-state or weak field regime \cite{viana2011}.
Very recently, the dynamics of a $\Lambda$-type atom interacting with non-overlapping ultrashort
laser train pulses  with a delta-function temporal dependence of the individual pulses  was studied in \cite{Ilnova}.

In this paper we study  the quantum interference between the excitation pathways in a
$\Lambda$-type atom interacting  with \textit{two trains of laser pulses} under the EIT conditions.
As specified before, under the conditions of EIT a medium could become transparent
for a probe laser by introducing a second laser:  coupling laser which is, usually,
 much stronger than the probe laser. We assume interaction  with two \textit{short} laser pulse trains,
with \textit{short} repetition periods, such that the atom does not have enough
time to completely relax between two consecutive single pulses and therefore
we expect some coherent accumulation effects during the propagation.
The paper is organized as follows. In Section \ref{mm}  we present the theoretical model.
 The time-dependent Schr\"odinger equation which describes the  interaction
of a  $\Lambda$-type atom with  \textit{a probe and a coupling laser pulse train} is numerically solved.
 Analytical results are presented for a particular case of laser trains with rectangular pulse shape.
In Section \ref{nr} we present illustrative results for the  interaction of two short
probe and  coupling laser pulse trains with a $\Lambda$-type atom,
for a coupling laser stronger than a probe laser.
We examine the effect of the parameters of the laser pulse trains such as pulse area,
laser detunings, and number of individual pulses, on the quantum interference in terms
of the upper excited state population. Finally, concluding remarks are given in Section \ref{su}.
Atomic units are used throughout this paper unless otherwise mentioned.

\section{Theoretical approach} \label{mm}

\subsection{The time-dependent probability amplitude equations}\label{am_eq}

It is well known that many atomic processes could be considered as involving only few levels and  using a three-level
$\Lambda $ model atom  provides a simple and  realistic description of the physical processes.
Our system under consideration is a three-level $\Lambda$-type atom interacting with two laser
fields: The states $| 1 \rangle$ and $| 2 \rangle$  are coupled by
 a probe laser $\bm {\mathcal{E}}_p(t)$ with photon energy $\omega_p$, while the states
$| 2 \rangle$ and $| 3 \rangle$  are connected by a coupling laser $\bm {\mathcal{E}}_c(t)$
with photon energy $\omega_c$.
The probe and coupling lasers are tunned near resonance with the respective  transitions,
and the transition between the states $| 1 \rangle$ and $| 3 \rangle$ is
dipole-forbidden.
Both the probe and coupling laser pulses  are considered as {\it laser pulse trains} and are described
by an electric field  which can be written as a sum of equally spaced identical electric fields

\bqn
\bm {\mathcal{E}}_{j}(t) =
\mathcal{E}_{0j} \;\bm {e_j}\; e^{i \omega_j t} \sum_{q=0}^{N-1} f_j(t-q T_j) e^{i q \phi} +c.c.,
\label{field_p}
\eqn
where  $ \mathcal{E}_{0j} $ is the peak amplitude of the electric field ($j=p$ or $c$ and hereafter),
 $\bm{e_j}$ is the polarization vector,  and $f_j(t) $ is the slowly varying envelope.
$T_j $ is the laser pulse repetition period that represents the time separation between
 two successive  single pulses,  $N$ is the number of individual pulses in the train pulse,
and $\phi$ is the phase shift  between the two successive pulses.
We assume both probe and coupling fields are linear polarized in the same direction,
 with the same pulse repetition period $T$ and the phase shift $\phi$ is neglected.
We consider that both probe and coupling lasers have the same time dependence,
with the temporal envelope of each single pulse in the train  described by  a Gaussian function

\bqn
f_j(t)= e^{-\pi \left( {t}/{\tau_j} \right)^2 } ,
\label{field_gauss}
\eqn
where $\tau_j$ is the temporal width of a single laser pulse.

By taking the Fourier transform of the electric field, equation (\ref{field_p}), with respect to time as

\bqn
 \mathcal{\widetilde E}_j(\omega) = \int_{-\infty} ^{+\infty} \mathcal{E}_j(t) e^{-i\omega t} d t,
\eqn
where the symbol $\sim$  denotes the  Fourier transform, we obtain, after simple algebra,

\bqn
 \mathcal{\widetilde E}_j(\omega) = \mathcal{ E}_{0j} \widetilde f_j(\omega - \omega_{j})
\frac{1- \exp{(i N \omega T)}}{1- \exp{(i \omega T)}}.
\label{field_p_spec}
\eqn

\noindent
The intensity of the laser pulse train in the frequency domain is calculated as

\bqn
 I _j(\omega) = |\mathcal{\widetilde E}_j(\omega)|^2 = { I}_{0j} (\omega)
\sin^2 \left( \frac{ N \omega T}{2} \right)
\sin^{-2} \left( \frac{\omega T}{2} \right) ,
\label{field_p_int}
\eqn
where $I_{0j}(\omega) =|2 \pi \mathcal{ E}_{0j} /T|^2 \left| \widetilde f_j(\omega - \omega_{j}) \right|^2  $
 represents the laser intensity of a single pulse.
In the limit of  $N \to \infty $ we obtain,  after a simple calculation,

\bqn
 I _j(\omega) ={ I}_{0j} (\omega)
\sum_{N=-\infty} ^{+\infty} \delta( \omega -\omega_{j N}),
\label{i_p_spec}
\eqn
where $\omega_{j N} =\omega_j+  N \omega_r$.
The frequency spectrum of the electric field given by equation (\ref{field_p})  consists of
a comb of $N^{th}$ laser modes that are separated by the repetition angular frequency
$\omega_r = 2\pi/T$ and centered at $\omega_j $.
The photon energy of the $N^{th}$ mode of laser pulse train is given by $\omega_j \pm  N \omega_r$.

In order to study the temporal evolution of the atomic system we derive the time dependent
Schr\"{o}dinger equation for the  interaction of a $\Lambda$-type atom with the probe and coupling lasers
in the standard framework  of rotating wave approximation, as long as the pulse duration
 is not very short (e.g. not in the fs time domain).
For simplicity reasons the Doppler broadening of the atomic levels is not included in the calculation.
By using the standard procedure, we obtain  the following set of time-dependent probability
amplitude equations for the interaction of the $\Lambda$-type atom, as illustrated in  Figure {\ref{fig1}},
 with the probe and coupling lasers given  by equation (\ref{field_p})

\bqna
 \frac{\partial}{\partial t} {u}_{1}(t) &=&   \; \frac{i}{2} {\Omega}_p (t) u_{2}(t),
\label{amp1} \\
 \frac{\partial}{\partial t} {u}_{2}(t)  &=&  -(i \delta_p + \gamma) u_{2}(t)
+ \; \frac{i}{2} {\Omega}_p (t) u_{1}(t) \nonumber \\
&&
+ \; \frac{i}{2} {\Omega}_c (t) u_{3}(t),
\label{amp2} \\
 \frac{\partial}{\partial t} {u}_{3}(t)  &=&  -i(\delta_p - \delta_c ) u_{3}(t)
+ \; \frac{i}{2} {\Omega}_c (t) u_{2}(t),
\label{amp3}
\eqna
where $ {u}_{k}(t)$  is the slowly varying probability amplitude of state
$| k \rangle$ ($k=1,2$, and $3$) that satisfies the initial conditions
$ {u}_{1}(t=-\infty)= 1 $ and ${u}_{k}(t=-\infty) =0 $ ($k=2$ and $3$).
Here $\Omega_{p} (t)$ and $ \Omega_{c} (t) $  are the one-photon Rabi frequency due
to the probe  and coupling lasers, which are defined as
$$
{\Omega}_{p}( t) = {\Omega}_{0p} \sum_{q=0}^{N-1} f_{p} ( t - q T ),
$$
$$
{\Omega}_{c} (t) = {\Omega}_{0c} \sum_{q=0}^{N-1} f_{c} ( t - q T ),
$$
\noindent
with the magnitudes  ${\Omega}_{0p}= d_{12} {\mathcal{E}}_{0p}$  and
${\Omega}_{0c}= d_{23} {\mathcal{E}}_{0c}$,
where $d_{12}$ is   the one-photon dipole moment for the transition between  states
$| 1 \rangle$  and  $| 2 \rangle$,
and $d_{23}$ is  the one-photon dipole moment for the transition between
the states $| 2 \rangle$  and  $| 3 \rangle$.
 $\delta_{p} $  and  $\delta_{c} $ are the
 one-photon detunings from the resonance of the probe and coupling lasers,  respectively,
and $\gamma $ is the spontaneous decay rate  of the upper excited state $| 2 \rangle$.
We introduce the temporal area of a single pulse, defined as  the integral
of the Rabi frequency over time

\bqn
\theta_j =\Omega_{0j}  \int_{-\infty}^\infty  f_{j} ( t )dt   \quad  \text{($j=p$ or $c$)}.
\eqn
For a laser pulse train with the Gaussian  temporal envelope given by equation (\ref{field_gauss}),
the area of a single pulse is obtained after a straightforward algebra as
$ \theta_p = \Omega_{0p} \tau_p $ for the probe laser and
$ \theta_c = \Omega_{0c} \tau_c $ for the coupling laser.

For laser pulses with arbitrary pulse envelopes and off-resonant photon energies,
when damping is included, a general analytic solution  of equations
(\ref{amp1})-(\ref{amp3}) cannot be derived.
Therefore we numerically integrate equations (\ref{amp1})-(\ref{amp3}) for
realistic  atomic and lasers parameters.
In order to understand the numerical results and the dynamics of the system
we analyze in the next subsection the particular case of two resonant laser pulse trains
with rectangular envelopes and the same temporal pulse areas
as those given by the Gaussian envelope equation (\ref{field_gauss}).

\subsection{Analytic solutions for resonant probe and coupling laser pulse trains}\label{ap1}

We analytically solve the set of three nonlinear coupled differential equations
 (\ref{amp1})-(\ref{amp3}) and derive an analytical solution for the population of
the upper excited state $| 2 \rangle$.
 We consider a probe and a coupling laser pulse train that are
 one-photon resonant with the $|1 \rangle-|2 \rangle$ and $|2 \rangle-|3 \rangle$ transitions,
respectively, $\delta_p=\delta_c=0$.

For simplicity,  we assume  laser pulse trains with  \textit{rectangular pulse envelopes} of
temporal width $\tau_p=\tau_c =\tau$, for both  probe and coupling lasers,
 with the following one-photon Rabi frequency

\begin{equation*}
\Omega_{j}(t)=
 \begin{cases}
 \Omega_{0j}, &   \quad  \text{for  \;   $ q T \leq    t \leq \tau + q T $}, \\
0, &    \quad \text{for  \;  $\tau +qT  \leq    t \leq \tau + (q+1)T$},
\end{cases}
\end{equation*}
where  $ 0 \leq    q \leq  N-1 $  and $j=p$ or $c$.
Since the atom interacts with two different laser fields we define the
total Rabi frequency   as the root mean square of the probe and
coupling Rabi frequencies
$\Omega_0=(\Omega_{0p}^{2} + \Omega_{0c}^{2})^{1/2}$.
It is worth mentioning that for resonant photon energies the
 transition probability depends on the temporal pulse area only
and is almost insensitive to the temporal envelope of the laser pulse \cite{Shore}.
Here the temporal areas of each single probe and coupling pulse are
$ \theta_p = \Omega_{0p} \tau_p $ and $ \theta_c = \Omega_{0c} \tau_c $,
respectively,
and are identical with those areas for probe and coupling
pulses with Gaussian temporal envelopes given by equation (\ref{field_gauss}).

Now for $   \gamma  < \Omega_0 $, which is often the case, equations (\ref{amp1})-(\ref{amp3})
can be easily solved for the first pulse in the train ($N=1$),  $ 0 \leq    t \leq  \tau  $,
and the population of the upper excited state at the end of the first  pulse reads as

\bqn
P_2^{(1)} =|u_{2}^{(1)}(\tau) |^2 =
\frac{\Omega_{0p}^2}{\Omega^{2}}   \; e^{-\gamma \tau} \; \sin^2 \left( \frac{ \theta}{2} \right),
\label{N1}
\eqn
 where $\theta = \Omega  \tau  $  represents the effective temporal
 pulse area of a single pair of probe and coupling pulses and
 $\Omega$ is the effective Rabi flopping frequency

\bqn
\Omega=(\Omega_0^2- \gamma^2)^{1/2}  \qquad  (    \gamma  < \Omega_0).
\nonumber
\eqn

Next, after a straightforward algebra we obtain  the population of the upper excited state
 $| 2 \rangle$ at the end of the  2$^{nd}$ ($N=2$), 3$^{rd}$ ($N=3$), and 4$^{th}$ ($N=4$)  pulse,
respectively,  calculated  for   $   \gamma  < \Omega_0$ as

\bqna
 P_2^{(2)} &=& \left| u_{2}(\tau +T) \right|^2 =
  4 \; P_2^{(1)}  e^{-\gamma T} \label{N2}  \\&\times&
 \left[ \cos  \left( \frac{ \theta}{2} \right)  \cosh \beta
+ \frac{\gamma}{\Omega} \sin  \left( \frac{ \theta}{2} \right) \sinh \beta \right]^2,
\nonumber \\
 P_2^{(3)}  &=& \left| u_{2}(\tau +2T) \right|^2=
 16 \;  P_2^{(1)}   e^{-2  \gamma T}  \label{N3} \\&\times&
 \left\{ \left[\cos   \left( \frac{ \theta}{2} \right)  \cosh\beta
 + \frac{\gamma}{\Omega} \sin  \left( \frac{ \theta}{2} \right) \sinh \beta \right]^2 -\frac{1}{4} \right\}^2,
 \nonumber\\
 P_2^{(4)}  &=&\left| u_{2}(\tau +3T) \right|^2 =
  64 \; P_2^{(1)}   e^{-3  \gamma T}
\cos^2   \left( \frac{ \theta}{2} \right)  \cosh ^2\beta \nonumber \\ &\times&
 \left[\cos^2   \left( \frac{ \theta}{2} \right)  \cosh ^2\beta -\frac{1}{2} \right]^2
    \qquad  (\gamma \ll\Omega_0),
\label{N4}
\eqna
where the argument of the hyperbolic functions is $\beta={\gamma \; (T-\tau)}/{2}$.
\noindent
After the interaction of the $\Lambda$-type atom with a train of
$N$ probe and coupling pulses, for $\gamma \ll\Omega_0$,
we obtain the following analytical formula for the population
of the upper excited state at the end of the $N^{th}$  pulse

\bqna
 P_2^{(N)} &=&
\left| u_{2}[\tau + (N-1)T] \right|^2 =
0.25 \; P_2^{(1)} \;e^{-(N-1) \gamma T}
\nonumber\\ &\times&
 \left| \left[ \cos^2 \left( \frac{\theta}{2} \right)  \cosh^2\beta
+i\sqrt{1- \cos^2 \left( \frac{\theta}{2} \right)  \cosh ^2\beta} \right]^N
 \right. \nonumber\\ &-&   \left.
\left[\cos^2 \left( \frac{ \theta}{2} \right)  \cosh^2\beta
-i\sqrt{1- \cos^2 \left( \frac{\theta}{2} \right)  \cosh^2\beta} \right]^N \right|^2
 \nonumber\\ &\times&
\left| 1 -\cos^2 \left( \frac{\theta}{2} \right)  \cosh^2\beta  \right|^{-1}
 \quad  (\gamma \ll \Omega_0).
\label{NN}
\eqna

\noindent
In the limit of short laser pulses  and small decay rate
compared with the pulse repetition frequency and the comb width
 ($\gamma  T  \ll 1 $ and $\gamma \tau \ll 1$), for  $\gamma \ll\Omega_0$,
equations (\ref{N1})-(\ref{N4}) further simplify as

\bqna
 &&  P_2^{(1)}\simeq
\frac{1}{1 +\Omega_{0c}^{2}/\Omega_{0p}^2}   \; \sin^2  \left( \frac{ \theta}{2} \right),
\label{N1s}\\
&&   P_2^{(2)} \simeq
\frac{e^{-\gamma T}}{1 +\Omega_{0c}^{2}/\Omega_{0p}^2}  \; \sin^2  \left(  \theta \right),
\label{N2s}\\
&&  P_2^{(3)}\simeq
\frac{e^{-2 \gamma  T}}{1 +\Omega_{0c}^{2}/\Omega_{0p}^2}  \;  \sin^2  \left( \frac{ 3 \theta}{2} \right),
\label{N3s}\\
&& P_2^{(4)} \simeq
\frac{e^{-3 \gamma T}}{1 +\Omega_{0c}^{2}/\Omega_{0p}^2}  \;  \sin^2  \left(  2  \theta \right) .
\label{N4s}
\eqna

\noindent
Furthermore, the population of the upper excited state $| 2 \rangle$
at the end of the $N^{th}$ pulse,
$P_2^{(N)}$, equation (\ref{NN}), could be easily expressed as

\bqna
P_2^{(N)}  &\simeq& P_2^{(1)} \; e^{- (N-1)\gamma T}  \;
\sin^2 \left( \frac{ N \theta}{2} \right)
\sin^{-2} \left( \frac{  \theta}{2} \right)
 \nonumber \\
&=&\frac{e^{- (N-1)\gamma T}}{1 +\Omega_{0c}^{2}/\Omega_{0p}^2}
 \; \sin^2 \left( \frac{ N \theta}{2} \right).
 \label{NNs}
\eqna

\noindent
It is worth pointing out that equation (\ref{NNs}) resembles equation (\ref{field_p_int}) that describes
the intensity of the laser pulse train in the frequency domain.

The simplified equations (\ref{N1s})-(\ref{NNs}) clearly show that the number of pulses,
the accumulated effective pulse area $N\theta$,
the repetition period, the spontaneous decay rate, and
the ratio of the probe and coupling Rabi frequencies
 play an important role in the accumulation effect induced by the pulse trains.
Population of the upper excited state oscillates sinusoidally with the effective pulse area $\theta$,
with an amplitude that quadratically increases with the ratio $\Omega_{0p}/\Omega_{0c}$
and exponentially decreases with $(N-1)\gamma T$,  and presents
minima for effective pulse areas $\theta_{min}= 2\pi m/N$ and
maxima for $\theta_{max}= (2 m+1)\pi/N$,  where $m \geq 0$ is an integer.
\noindent
For the case of a two-level system, by turning off the coupling laser ($\Omega_{0c} =0$) and
neglecting the spontaneous decay rate ($\gamma \simeq 0$), the population of the upper excited state
given by equation (\ref{NNs}) reduces to the  corresponding expression
derived in \cite{Vitanov} for the exact resonance case of a non-decaying two-level  atom.
From our knowledge there are other analytical results derived for a single laser pulse train
with large detunings \cite{Nakajima2008}, or based on an iterative formula in \cite{aumiler},
or for two laser pulse trains  with pulse envelope described by  a delta-function \cite{Ilnova}.

\noindent
In the limit of weak excitations  for quite small effective pulse areas,
 the population of the excited state at the end of the $N^{th}$ pulse
given in equation (\ref{NNs}) becomes

\bqn
P_2^{(N)} \simeq  P_2^{(1)} \;  N^2  \; e^{- (N-1)\gamma T},
\label{NNl}
\eqn
that shows how the coherent accumulation effects in the excited state population develop
from one single pulse  $(N=1)$ to a train of $N$ pulses. That is, compared with the population
 of a single pair of pulses ($N=1$), the population
$P_2^{(N)} $ increases  quadratically with  the number of pulses $N$ but
 is  exponentially damped by the factor  $e^{-(N-1) \gamma T}$.

\section{Numerical Results} \label{nr}

In this section we present and discuss representative results for the interaction of a
$\Lambda$-type atom with short probe and coupling laser pulse trains
 in the framework of EIT phenomenon where the coupling
laser is considered stronger than the probe laser.
The probe and coupling lasers are described by train pulses given by equation ({\ref{field_p}}),
 with an envelope of each single pulse  assumed to be a Gaussian function denoted by
equation ({\ref{field_gauss}}).
We consider that both pulse duration $\tau$ and  repetition period $T$
are shorter than the lifetime of the upper state of the $\Lambda$-type atom.
Because  the repetition period $T$ is  short compared to the upper state lifetime,
such that the atom does not have enough time to completely decay between
two consecutive pulses,  we expect some accumulation of atomic coherences
from one pulse to the next one in the train \cite{Felinto2001,Stowe},
which is consistent to analytical results equations ({\ref{NNs}}) and  ({\ref{NNl}}).
This implies, unless the excitation is not extremely small, that each single pulse in the
 laser train finds the atom in a different initial state than the previous pulse.
Furthermore, if for a single pulse the EIT conditions are not initially fulfilled
 (such that the coupling pulse is not much stronger than the probe pulse)
and the probe laser is absorbed, then for a short \textit{laser pulse train} with realistic
 repetition rate, we  expect that the EIT conditions could be reached for a large
 number of pulses $N$ due to atomic coherence accumulations from one pulse to the next one.

We numerically integrate the set of three nonlinear coupled partial differential equations
(\ref{amp1})-(\ref{amp3}) and calculate the population of the state $| k \rangle$  as
 $P_k(t) =|u_k(t) |^2$,  where $k=1, 2$, and $3$.
The atomic coherence between the states $| 1 \rangle$ and $| 2 \rangle$ is defined by
 the product $u_1(t) u_2^*(t) $,  where the real part, Re$ [u_1(t) u_2^*(t) ]$,  denotes
 the refraction index and the imaginary part, -Im$ [u_1(t) u_2^*(t) ]$,  denotes the
absorption coefficient if it is negative or amplification if it is positive, respectively.
Whenever it is possible the numerical and analytical results are compared.

\subsection{Quantum interference for different  probe pulse areas}

It is well known that Rabi oscillations are used as a tool to coherent control the quantum
dynamics of atomic systems, to measure the pulse area, and the excited state population \cite{Allen}.
In this subsection we  present numerical and analytical results for small,
 moderate, and large coupling  pulse area, for resonant  probe and coupling laser trains.

In  Figures {\ref{fig1}}a-{\ref{fig1}}c we plot the population of the upper excited state
 $P_2^{(N)}$ at the end of the laser pulse as a function of the probe pulse area
$\theta_{p}=\Omega_{0p}\tau$ for the first ($N=1$), $2^{nd}$ ($N=2$), $4^{th}$  ($N=4$),
and $10^{th}$ ($N=10$) pulse in the train.
The laser parameters are  $\Omega_{0c} =0.1 $ THz (Figure {\ref{fig1}}a),
$1 $ THz (Figure {\ref{fig1}}b), and $10 $ THz (Figure {\ref{fig1}}c),  pulse duration
 $\tau=1$ ps,  repetition period  $T=1$ ns, and  laser detunings  $\delta_p=\delta_c= 0 $.
The spontaneous decay rate from the upper state is  assumed  everywhere in this paper
as $\gamma= 70 $ MHz, that corresponds to a 14 ns lifetime.
 For  Figure {\ref{fig1} it gives the product $\gamma T= 0.07$.
As expected, at resonance, the excited state population exhibits periodic Rabi oscillations
and we found out that the simplified analytical results,
 derived for laser pulse trains with rectangular pulse envelopes,
given  by equations (\ref{N1s})-(\ref{NNs}) and indicated by  solid circles in Figure {\ref{fig1}},
are in very good agreement with the numerical results.

\noindent
For a small coupling pulse area  $\theta_{c}=\Omega_{0c}\tau =0.1$ the population
$P_2^{(N)} $, shown in Figure {\ref{fig1}}a, oscillates sinusoidally with the probe pulse
area according to the analytical solutions (\ref{N1s})-(\ref{NNs}).
In Figure {\ref{fig1}}a the number of Rabi oscillations of $P_2^{(N)}$  induced by the train pulses
increases with the number of pulses $N$
  and population is exponentially damped by the factor $e^{-(N-1) \gamma T}$,
as we  notice from equation (\ref{NNs}).
The minima of the population $P_2^{(N)}$ correspond to destructive interference
between the $|1 \rangle-|2 \rangle$ and $|2 \rangle-|3 \rangle$ transitions,
 which occur for probe pulses with an area of $\theta_{p} \simeq  2 \pi m/N$,
 where $m$ is an integer.  The value of $m$ is such that $m > m_0$, with  $m_0= [ N \theta_{c}/2\pi ]$,
where $[x]$ denotes the integer part of $x$.
By increasing in Figures {\ref{fig1}}b and {\ref{fig1}c
the area of the coupling pulse to $ \theta_{c} $ = 1 and 10,
 the excited state population is furthermore attenuated by the Rabi frequencies
ratio ${\Omega_{0p}^2}/{\Omega_{0c}^2 }$  [see equation (\ref{NNs})] and
the magnitude of the  peaks of $P_2^{(N)}$ located at small values of the probe pulse area
  decreases as the number of pulses in the train increases.
The minima of the population $P_2^{(N)}$ in Figures \ref{fig1}b and \ref{fig1}c, are related to
the pair of probe and coupling pulses with an effective Rabi frequency area of
$ \theta=\Omega \tau = 2  \pi m /N$,  with $m >m_0$.

\noindent
Next, in Figure {\ref{fig2}} we present similar results to those shown in  Figure {\ref{fig1}}
but for a longer laser repetition period of $10$ ns that corresponds to $\gamma T= 0.7 $.
The numerical results for the population of the upper excited state  and  the analytical
 results (shown by the solid circles) given by equations (\ref{N1})-(\ref{N2}) and (\ref{N4})-(\ref{NN}),
   for $ N=1,2, 4$, and $10$, are in very good agreement.
Of course, compared with a $1$ ns laser repetition period ($\gamma T= 0.07 $)
 we expect the upper state population $P_2^{(N)}$ to be more damped as the
number of  pulses in the train increases,  and we notice in
Figures {\ref{fig2}}b and {\ref{fig2}}c that after  interaction with
 $10$ pulses  population $P_2^{(10)}$ is strongly suppressed and
significant Rabi flopping occurs only for a pair of probe and coupling
pulses with an effective Rabi frequency area $\theta= 2 \pi m$,  with $m>m_0$.
This combination of probe and coupling laser pulses is the equivalent of the so called $2\pi$-pulse
 for a two-level atom that transfers the population from the initial state $|1\rangle$ to  the
 excited state $|2\rangle$ and back again to $|1\rangle$, while for a three-level atom the population
is transferred to a superposition of the two lower states  $|1\rangle$ and  $|3\rangle$, creating a  dark state.
Actually, the fact that the first peak of the upper state population
 almost vanished in Figures {\ref{fig2}}b and {\ref{fig2}}c at small probe pulse areas,
is one of the consequences of the EIT as well as CTP effects, where the population is trapped
 between the states $| 1 \rangle$ and $| 3 \rangle$, while the population of the upper
 excited state  $| 2 \rangle$ is negligible.
In the next subsection we resume and discuss our results for the population dynamics.

\noindent
In agreement with the theoretical results equations (\ref{N1})-(\ref{NN}),
for weak and moderate coupling pulse
 areas, $\theta_{c} \le 1 $ in Figures {\ref{fig2}}a and {\ref{fig2}}b,
 the common minima of the populations $P_2^{(N)}$ are located around the probe pulse area of
 $\theta_{p} \simeq 2  \pi m$,  while for larger coupling pulse area $\theta_{c} =10$,
 in Figure \ref{fig2}c,  the  common minima are located around the values of the probe pulse area
$\theta_{p} =(2  \pi m -\theta_{c}^2  )^{1/2}$, $m >m_0$.
What we learned for Figures {\ref{fig1}} and {\ref{fig2}} is that
coherent accumulation of excitation decreases for larger coupling pulse areas
 and is exponentially attenuated by a term which is proportional to the
 laser repetition period and number of pulses.
Depending on the repetition rate  the EIT effect occurs for a coupling pulse area
larger than some critical value.

\subsection{Quantum interference for different  probe pulse detunings}

Now, it is interesting  for spectroscopic investigations to study the upper
excited state population as a function of the probe laser detuning for different
 coupling pulse areas and  repetition periods.
We show in  Figures {\ref{fig3}}a-{\ref{fig3}}c the population of the  excited state
$P_2^{(N)}$  versus the detuning of the probe laser pulse train.
The number of individual pulses of the probe and coupling trains are
$N = 1,2,4,10$, and $40$, respectively, from top to bottom.
The laser parameters are  $\Omega_{0p} =0.04 $ THz and $\Omega_{0c} =0.1 $ THz
(Figure {\ref{fig3}}a), 1 THz (Figure {\ref{fig3}}b), and 10 THz (Figure {\ref{fig3}}c).
Both laser pulse trains have a Gaussian temporal profile with a duration
of each single pulse of $ 1$ ps and repetition period  of 1 ns.
For simplicity the coupling laser is considered at resonance, $\delta_c=0$.
It is clear from Figure {\ref{fig3}} that the excited state population changes
with the detuning of the probe laser and its periodic structure represent
the optical Ramsey fringes \cite{Ramsey1950}-\cite{Thomas1987}.
The resonances occur whenever the  probe laser detuning is a multiple of the repetition
angular frequency $\omega_r$ and the oscillation period of population $P_2^{(N)}$ is  exactly
the period of the frequency spectrum of the laser pulse train [see equation (\ref{field_p_int})].
The detailed shape of the central Ramsey fringe and the evolution of the resonances with $N$
 is presented in detail in Figures {\ref{fig4}}a-{\ref{fig4}}c for the same parameters
as in Figures {\ref{fig3}}a-{\ref{fig3}}c.

In Figures {\ref{fig5}} and   {\ref{fig6}}  are presented similar results as in
 Figures {\ref{fig3}} and  {\ref{fig4}},  but for a longer repetition period of 10 ns.
From Figures {\ref{fig3}}-{\ref{fig6}} it is clear that the height of the peaks in
the population profile  increases with the number of pulses in the train $N$,
while the width of the population profile  becomes narrower as $N$ increases.
Depending on the values of the coupling pulse area, we notice three different regimes
for the variation of the population  $P_2^{(N)}$   with probe laser detuning:
\begin{itemize}
\item[(i)]
 The regime of small coupling pulse area, $\theta_{c} =0.1 $,  Figures {\ref{fig4}}a and
 {\ref{fig6}}a, where the population of the excited state increases
 (coherent accumulates) with the number of
pulses $N$ for probe laser detunings $\delta_p=m \omega_r$, where $m$ is an integer.
Only after excitation with more than $20$ pulses a small dip appears in the middle of the
population profile due to the destructive quantum interference between the two excitation
pathways: $|1\rangle $-$|2\rangle $ and $|2\rangle $-$|3\rangle $.
\item[(ii)] The regime of moderate coupling pulse area, $ \theta_{c} =1 $, shown
in Figures {\ref{fig4}}b and  {\ref{fig6}}b,
where  the EIT effect occurs  and  the population of the excited state
 $P_2^{(N)}$  exhibits two distinct peaks as $N$ increases, and
 is negligible at resonance $\delta_p=m \omega_r$.
\item[(iii)] The regime of large coupling pulse area, $ \theta_{c} =10 $,  that is presented
in Figures {\ref{fig4}}c and   {\ref{fig6}}c,  where  the two peaks of $P_2^{(N)}$
are clearly separated by a gap in the frequency domain. This effect is known as the
Autler-Townes splitting \cite{autler}.
\end{itemize}
\noindent
The results presented in Figures {\ref{fig4}} and {\ref{fig6}} show that the
coherent accumulation also plays an important role for small and moderate areas of
 the coupling pulse at off-resonant probe laser detunings.
Our results  are consistent with the results obtained
for a closed $\Lambda$-type system interacting with a femtosecond laser pulse
 train and a  CW laser  by an  iterative numerical  method \cite{Soares2010}.

Next, in order to understand the role played by the number of pulses in the trains,
 populations $P_k(t) =|u_k(t) |^2$,  ($k=1, 2$, and $3$) and absorption,
 -Im$ [u_1(t) u_2^*(t)]$, are plotted  as a function of time for a repetition
 period of $1$ ns ($\gamma T=0.07$) in  Figures {\ref{fig7}}a-{\ref{fig7}}d, and
 $10$ ns ($\gamma  T=0.7$) in Figures {\ref{fig7}}e-{\ref{fig7}}h.
The probe and coupling laser parameters are $\Omega_{0p} =0.04 $ THz and
$\Omega_{0c} =0.2 $ THz  (red solid line), 0.5  THz (green dashed line),
and 1 THz (blue dot-dashed line), $\tau = 1$ ps,  and $\delta_p=\delta_c=0$.
Clearly,   Figures {\ref{fig7}}a-{\ref{fig7}}d and  Figures {\ref{fig7}}e-{\ref{fig7}}h
  show different dynamics of the populations and absorption for  short and large repetition period.
For small coupling pulse area $\theta_{c}=0.2 $ and  $T=1$  ns,  in  Figure {\ref{fig7}}b,
the population $P_2$ increases with time (the excitation accumulates
since there is less time for atom to relax between two consecutive pulses),
 it reaches a maximum value for $t\simeq 13 T$, and  after that it decreases to negligible values.
The absorption  (Figure {\ref{fig7}}d) does not take negligible values and
 there is a small population transfer from state $|1\rangle$ to  state $|3\rangle$.
For larger coupling pulse areas  $ \theta_{c} =0.5 $ and $1$  the population
$P_2$ oscillates sinusoidally in time.
For a longer repetition period $T=10$  ns (Figures {\ref{fig7}}e-{\ref{fig7}}h)
both  population $P_2$ (Figure {\ref{fig7}}f) and absorption  (Figure {\ref{fig7}}h)
accumulate very fast during the first few pulses and then start to decrease.
In comparison with the repetition period   of $1$ ns
population $P_2$ and  absorption present a strong oscillatory behavior  from one pulse
to the next one in the train,  a sawtooth profile  that describes successive
excitations of the upper level followed by its spontaneous decay \cite{Felinto2004}.
 For larger coupling pulse areas $\theta_{c} =0.5 $ and $ 1 $,
after the EIT and AT regime occur for $t$ larger than $30 T$  and $10 T$, respectively,
 the populations  $P_1$ and $P_3$ reach stationary
 values that do not change with the number of pulses $N$,
while the excited state population $P_2$ and absorption
are negligible (in Figures {\ref{fig7}}e-{\ref{fig7}}h).
This temporal   dynamics explains the evolution of the EIT and AT resonance of
$P_2$ in  Figures {\ref{fig3}}-{\ref{fig6}}.
The analytical results calculated at the end of each pulse given by equation (\ref{NN})
(indicated by solid circles in Figures {\ref{fig7}}b and {\ref{fig7}}f),
are in good agreement with the numerical results.

\section{Summary} \label{su}

In this paper we have studied the quantum interference between the excitation pathways in a
 three-level $\Lambda$-type atom  interacting  with  short probe and coupling  laser pulse trains,
beyond the steady state approximation, under EIT conditions.
We  have investigated the modification induced by the laser pulse trains in a lambda-type atom in terms of
upper excited state population for different pulse areas and different detunings.
We have numerically integrated the  probability amplitude equations that describe the
 interaction of the $\Lambda$-type atom with  the two  laser pulse trains.
For  resonant laser pulse trains with a rectangular temporal profile we have derived analytical formulas
 for the population of the upper excited state at the end of the pulse.
For the atomic and laser parameters used in the present paper
 we obtain a very good agreement between the analytical results
(calculated for rectangular pulse envelopes) and the numerical results
 (calculated for Gaussian pulse envelopes with identical pulse areas as for rectangular shape).
We have discussed the dynamics of the upper state  population and presented
numerical and analytical results for small, moderate, and large coupling  pulse areas for resonant
 probe and coupling laser trains.
We have showed that we can control the interaction of a $\Lambda$-type atom with two laser pulse trains
 under the EIT conditions, for small probe pulse area while the area of the coupling pulse is moderate,
 by manipulating certain parameters of the lasers such that:
 Rabi frequencies,  pulse repetition period, number of individual pulses, and  detunings.

\begin{acknowledgement}
The work by G.B. was  supported by research programs Laplas 3 PN 09 39N and Ro-Fair,
from the Ministry of Education and Research of Romania.
\end{acknowledgement}

\clearpage

\begin{figure}
\vspace{6mm}
\centering
\includegraphics[width=3.5in,angle=0]{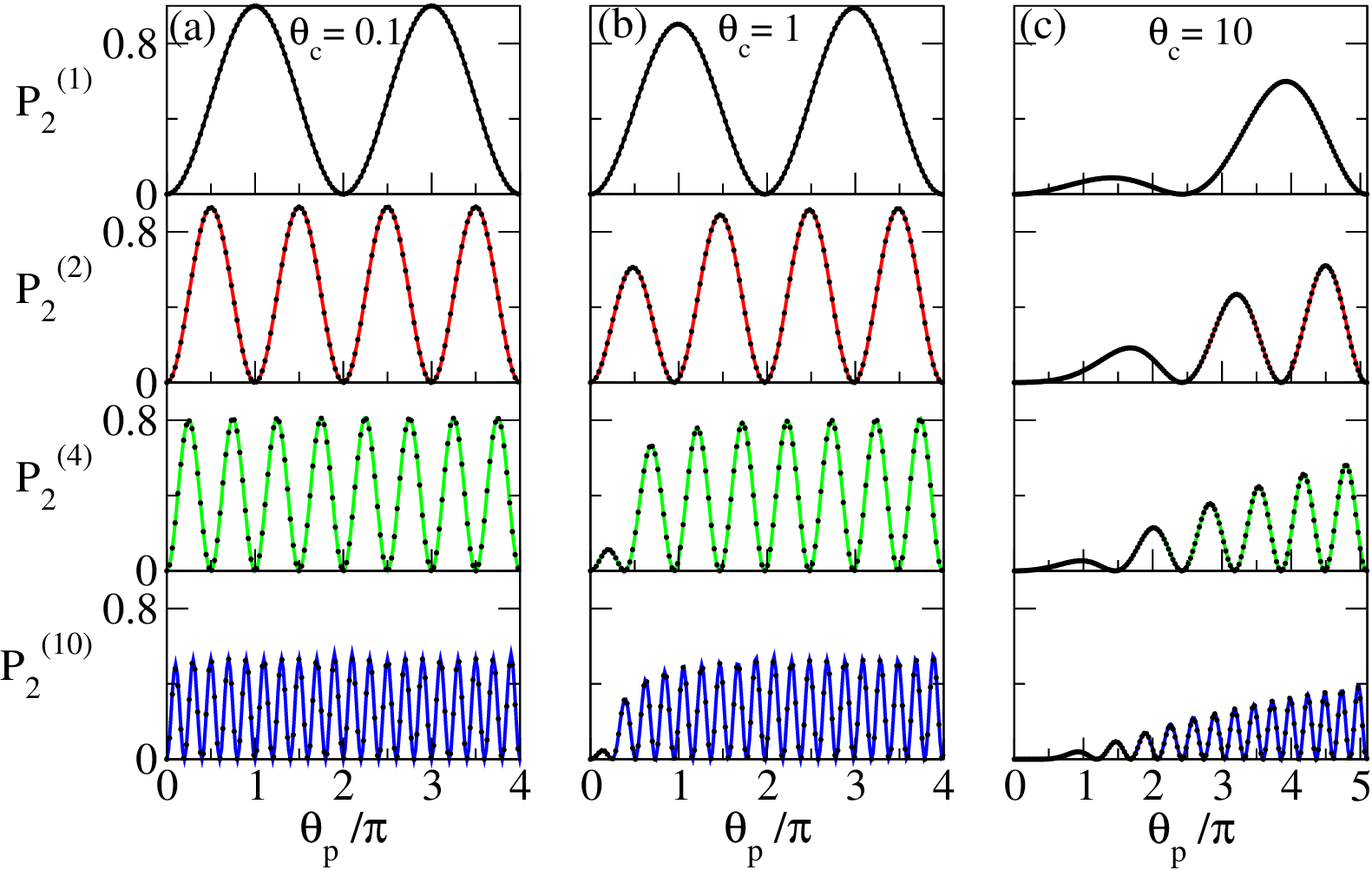}
\caption{(Color online)
Population of the upper excited state   $P_2^{(N)}$  as a function of the probe laser area.
The solid lines show the numerical results while the solid circles shows the analytical results.
The number of individual pulses in the trains  are $N =1, 2, 4,$ and $10$ (from top to bottom).
The coupling laser area $\theta_{c}  $  is  $0.1 $  in (a), $1$ in (b), and $10$ in (c).
Both laser pulse trains have a Gaussian temporal envelope with the duration $\tau = 1$ ps,
 repetition period  $T =1 $ ns, and  $\delta_p=\delta_c=0$.
The spontaneous decay rate is $\gamma=70 $ MHz which results in the  value $\gamma T=0.07 $.
}
\label{fig1}
\end{figure}

\begin{figure}
\vspace{6mm}
\centering
\includegraphics[width=3.5in]{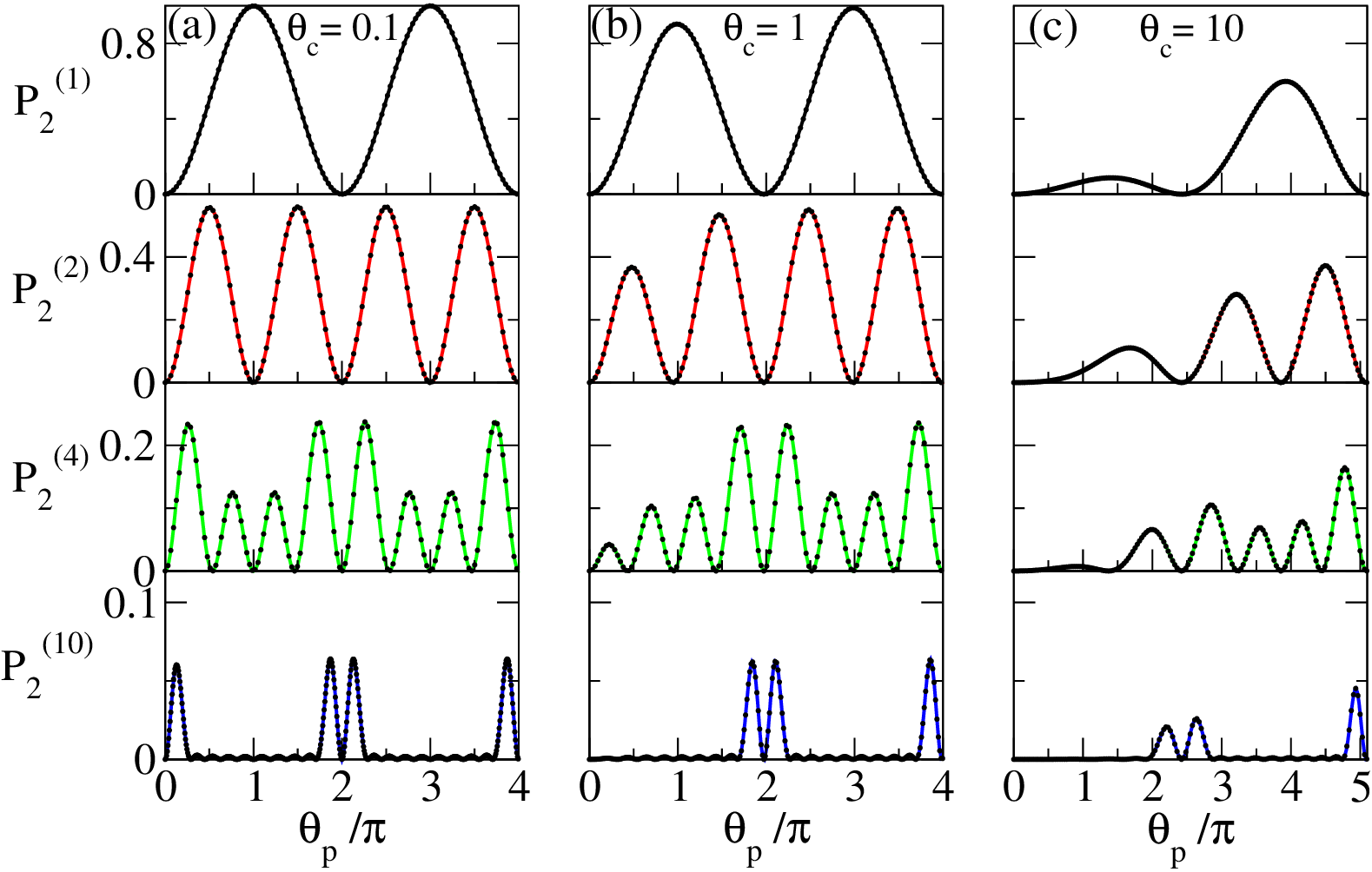}
\caption{(Color online)
Similar results to those in Figure \ref{fig1} but for  a longer  repetition period  $T =10 $ ns.
}
\label{fig2}
\end{figure}

\begin{figure}
\vspace{6mm}
\centering
\includegraphics[width=3.4in,angle=0]{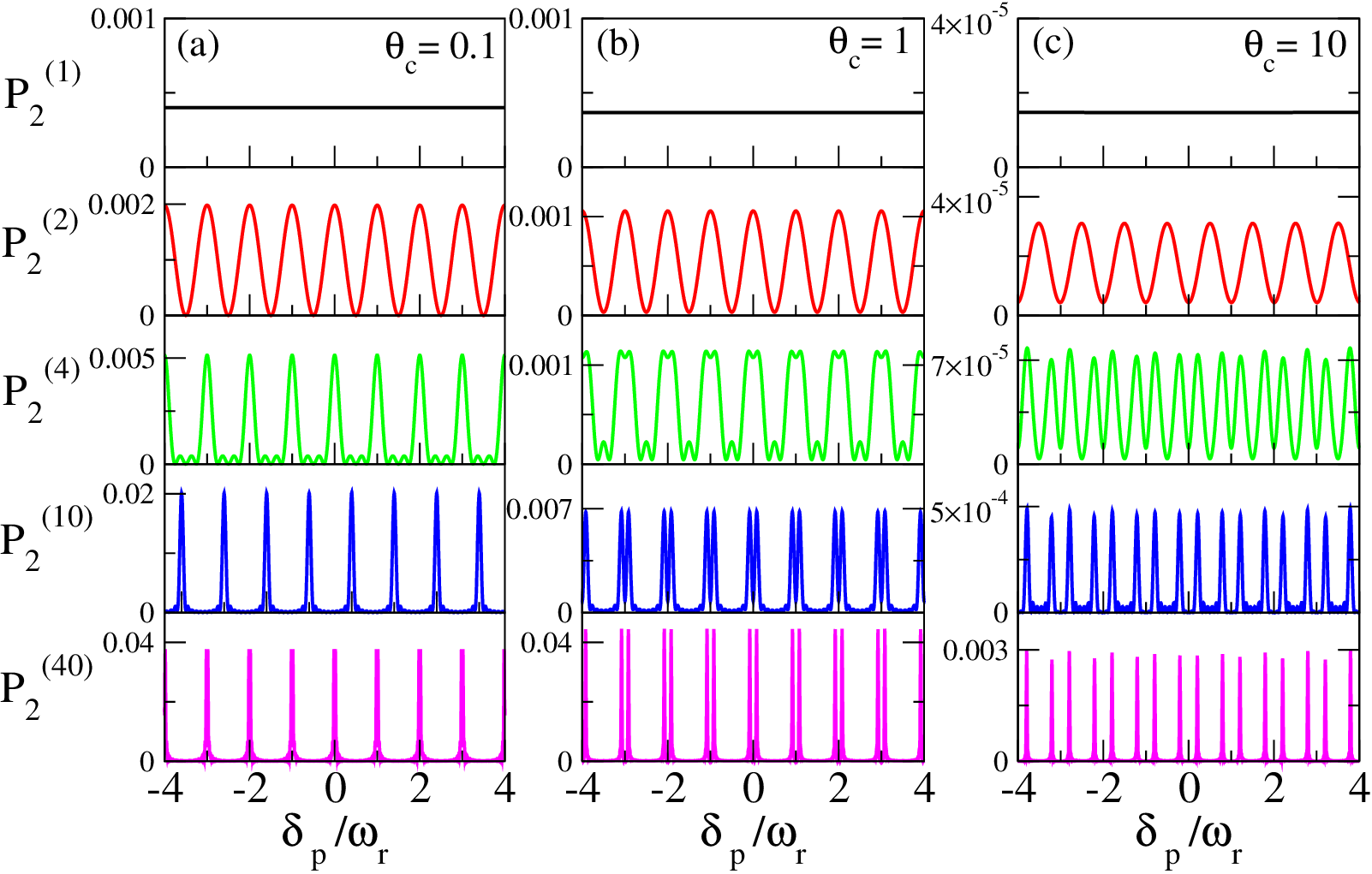}
\caption{(Color online)
Population of the upper excited state  $P_2^{(N)}$  as a function of the probe laser detuning.
The number of individual pulses in the trains  are  $N =1, 2,4,10,$ and $40$ (from top to bottom).
The probe laser area is $\theta_{p} =0.04 $  and the coupling laser area $\theta_{c}  $
is $0.1 $  in (a), 1 in (b) and 10 in (c), from top to bottom.
Both laser pulse trains have a Gaussian temporal envelope with the duration  $\tau = 1$ ps,
 the repetition period  $T =1 $ ns, and the coupling laser detuning is $\delta_c=0$.}
\label{fig3}
\end{figure}

\begin{figure}
\vspace{6mm}
\centering
\includegraphics[width=3.5in,angle=0]{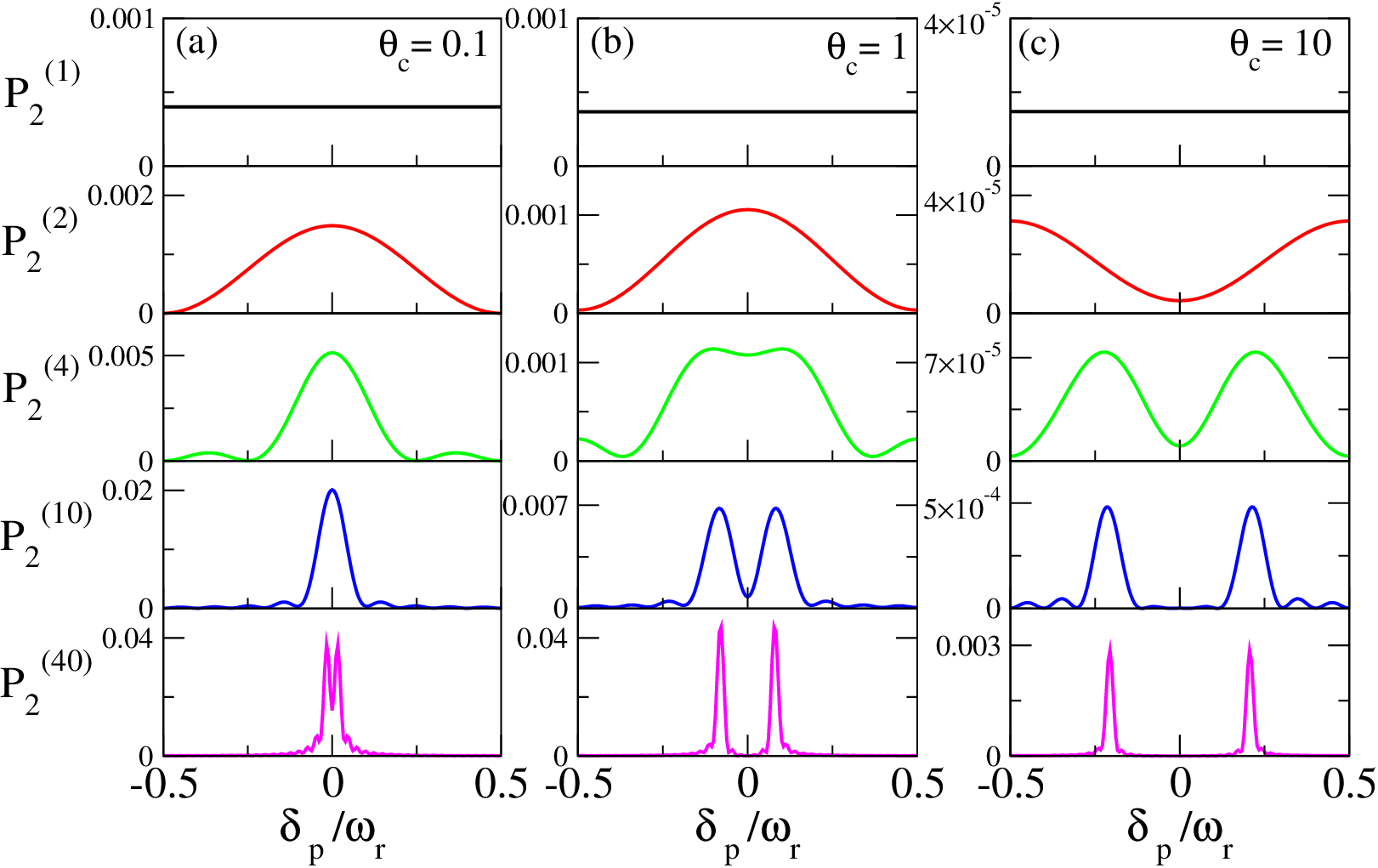}
\caption{(Color online)
The same  results as those in Figure \ref{fig3}, but detailed for a small detuning range
around $-  \omega_r/2   \leq \delta_p  \leq \omega_r/2$, representing the central Ramsey fringe.}
\label{fig4}
\end{figure}

\begin{figure}
\vspace{6mm}
\centering
\includegraphics[width=3.4in,angle=0]{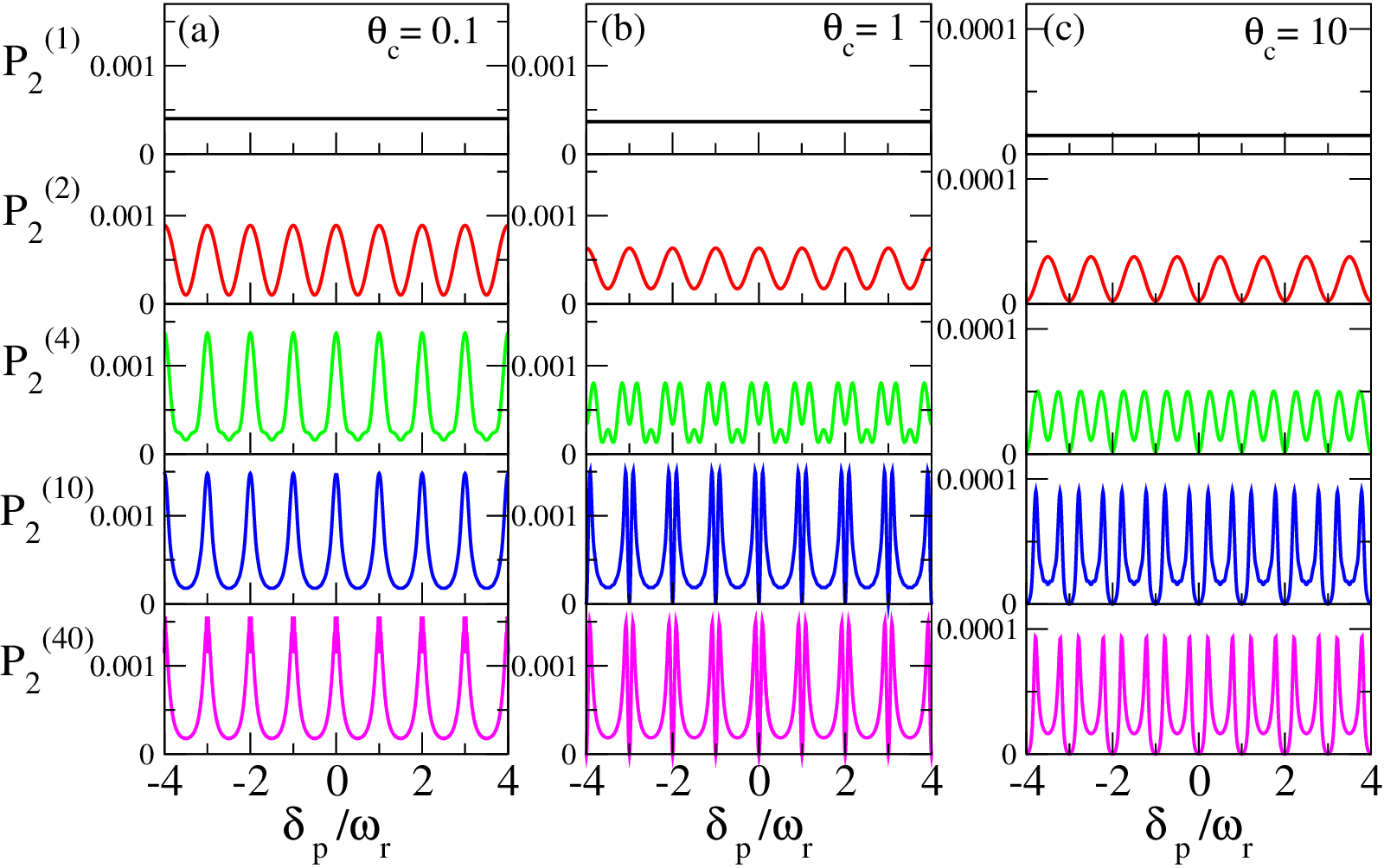}
\caption{(Color online)
Similar results to those in Figure \ref{fig3} but for  a repetition period  $T =10 $ ns.}
\label{fig5}
\end{figure}

\begin{figure}
\vspace{6mm}
\centering
\includegraphics[width=3.5in,angle=0]{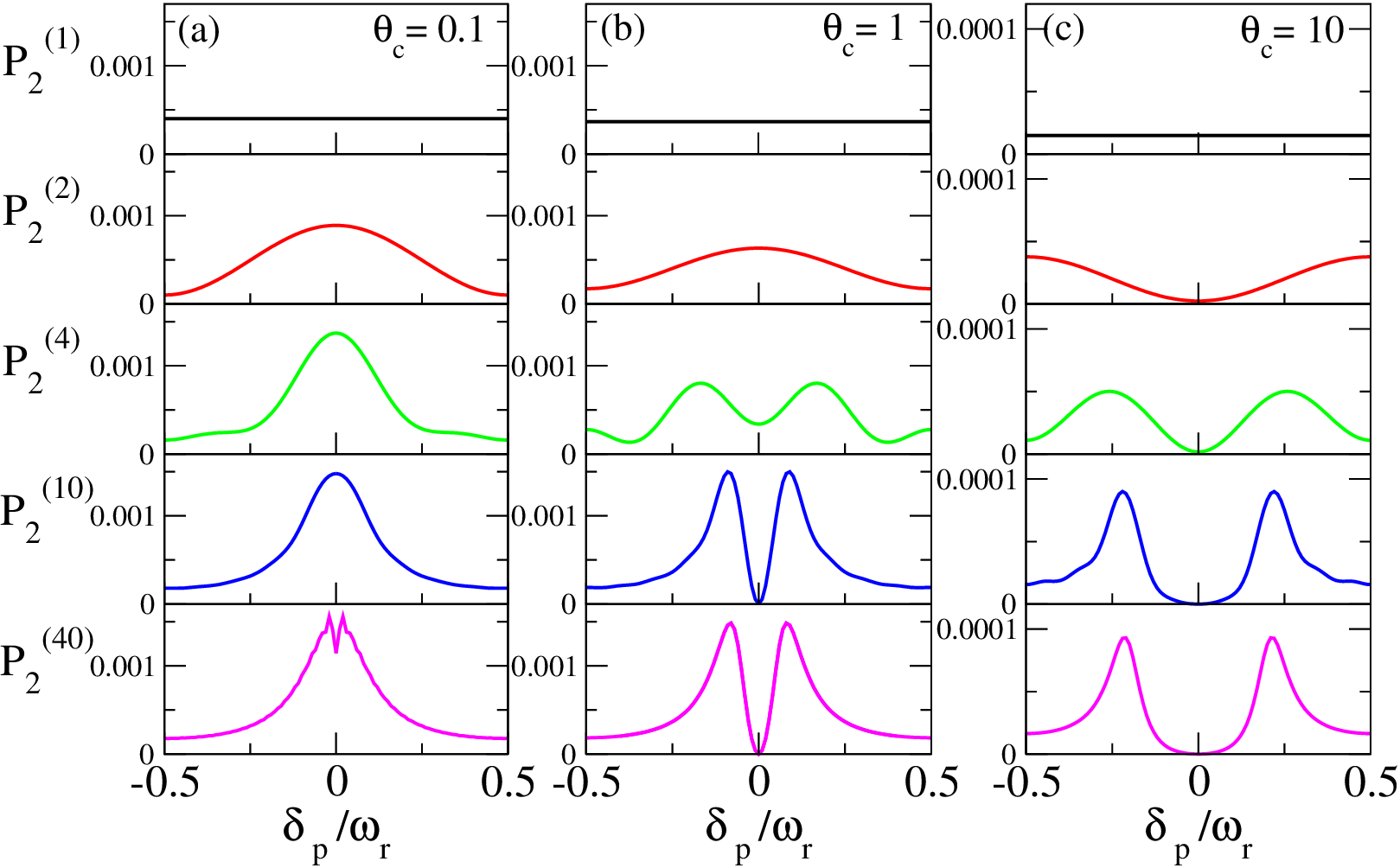}
\caption{(Color online)
The same results as those in Figure \ref{fig5}, but detailed for a small detuning range
around $- \omega_r/2   \leq \delta_p   \leq \omega_r/2$, representing the central Ramsey fringe.}
\label{fig6}
\end{figure}

\begin{figure}
\vspace{6mm}
\centering
\includegraphics[width=3.5in,angle=0]{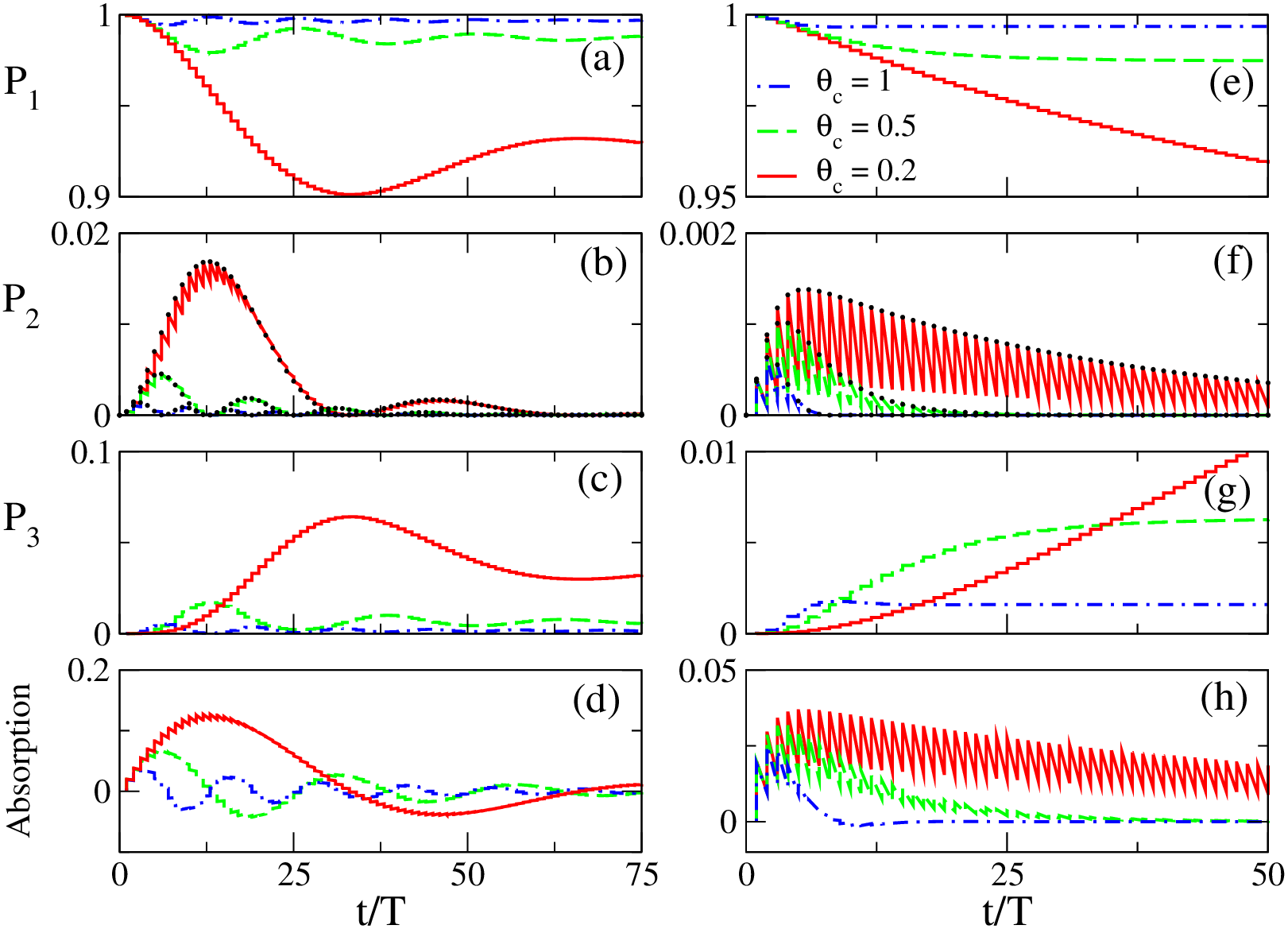}
\caption{(Color online)
Populations  $P_1$, $P_2$, and $P_3$ and absorption as a function of time
for a repetition period  of $1 $ ns in (a)-(d) and $10 $ ns in (e)-(h).
The probe laser area is $\theta_{p} =0.04 $  and the coupling laser area is
$ \theta_{c}  =0.2 $ (red solid line), $0.5 $ (green dashed line), and $1 $ (blue dot-dashed line).
The solid  circles in  (b) and (f) represents the analytical results calculated at the end of each pulse.
Both laser pulse trains have a Gaussian temporal envelope with the duration $\tau = 1$ ps
 and  detunings $\delta_p=\delta_c=0$.
}
\label{fig7}
\end{figure}

\end{document}